\documentstyle[preprint,aps]{revtex}
\def\ra{\rightarrow}
\def\be{\begin{equation}}
\def\ee{\end{equation}}
\def\bea{\begin{eqnarray}}
\def\eea{\end{eqnarray}}
\def\lln{\left<}
\def\rln{\right>}
\def\gam{\gamma}
\def\lb{\Lambda}
\def\lbb{\Lambda_b}

\def\gff{{G_f \over {\sqrt{2}}}}

\def\prl{Phys. Rev. Lett.~}
\def\pr{Phys. Rev.~}
\def\pl{Phys. Lett.~}
\def\np{Nucl. Phys.~}
\def\prl{Phys. Rev. Lett.~}

\begin{document}
\title{T-violating triple product asymmetries in $\lbb \ra \lb \pi^+
\pi^-$ decay}
\author{S. Arunagiri\footnote{arun@phys.nthu.edu.tw} and C. Q.
Geng\footnote{geng@phys.nthu.edu.tw}}
\address{Department of Physics, National Tsing Hua University,\\
Hsinchu, Taiwan 300, ROC}
\date{\today}
\maketitle
\begin{abstract}
We study $\lbb \ra \lb \pi^+ \pi^-$. The branching fraction is
predicted to be about 4.4 $\times 10^{-8}$. Within the standard
model, we compute $T$-odd triple product asymmetries to be about
1.4, 4.3, 6.5 and 7.2\% respectively due to $\vec s_{\lbb,\lb}
\cdot (\vec p_{\lb} \times \vec p_\pi)$ and $\vec p_{\Lambda,\pi}
\cdot (\vec s_\lb \times \vec s_{\lbb})$.
\end{abstract}
\vskip1.0in
PACS number(s): 11.30.Er, 13.30.Eg, 14.20.Mr
\newpage
In this paper, we consider $\lbb \ra \lb \pi^+ \pi^-$ decay and
study $T$-odd triple product correlations (TPC's). As has been
discussed in \cite{gv,bk,dl,sa}, the presence of a nonzero TPC of the 
form 
$\vec v_1 \cdot (\vec v_2 \times \vec v_3)$, where $\vec v_i$'s are 
spin 
or momentum, is given by
\be 
A_T = {
\Gamma (\vec v_1 \cdot (\vec v_2 \times \vec v_3) > 0) -
\Gamma (\vec v_1 \cdot (\vec v_2 \times \vec v_3) < 0) \over {
\Gamma (\vec v_1 \cdot (\vec v_2 \times \vec v_3) > 0) +
\Gamma (\vec v_1 \cdot (\vec v_2 \times \vec v_3) < 0)}}
\ee
where $\Gamma $ is the decay rate in question. In order to see that the
the TPC is indeed caused by weak phase, it is expressed in comparison 
with 
that of the  corresponding conjucate mode, $\bar A_T$, as triple 
product 
asymmetry (TPA)
\bea {\cal {A}}_T &=& {1 \over
2} (A_T - \bar A_T) \nonumber \\ &\sim& \sin \phi \cos \delta \eea
$\phi (\delta)$ is weak (strong) phase. 
Thus, TPA is maximal in the vanishing limit of the strong phase.
In $\lbb \ra p P(V)$ where
$P(V)$ is pseudoscalar (vector) meson, T-odd triple correlations
give rise to an asymmetry of about $O(10\%)$ in certain decays
while in some at $O(1\%)$ \cite{dl}. However, in beyond standard
model scenarios the TPA is about 50\% or so \cite{dl0,dl1}. Thus
$\lbb$ decays provides better opportunity to study CP/T violating
effects. For instance, $\lbb$-spin and the final state baryon spin
are obervable in terms of which physics of the standard model and
beyond it can be investigated.
In the two-body decay mode, i. e., $\lbb \ra p P$, TPC's
necessarily involve the polarisation of both baryons. Whereas in
three-body final state such as one we consider, TPC's are
constituted by either of the baryon polarisations, (besides cases
involving both). This is more feasible experimentally, despite the
size being relatively smaller, than the one with two
polarisations. On the other hand,  with the most of the $b$ quark
polarisation being carried by the $\lbb$ baryon, polarisation of
final state baryon alone can be considered in the search for CP
violating effects. Hopefully, at the future colliders BTeV, LHC-b,
production of about $10^{10}$ $\lbb \bar \lbb$ pairs would
facilitate such studies.
In this work, we describe the decay using factorisation and obtain the
branching ratio. Then, we look for the following TPC's:
$\vec s_{\lbb,\lb} \cdot (\vec p_{\lb} \times \vec p_\pi)$
and
$\vec p_{\Lambda,\pi} \cdot (\vec s_\lb \times \vec s_{\lbb})$.
The effective Hamiltonion for $b \ra u \bar u s$ which underlies the
decay $\lbb \ra \lb \pi^+ \pi^-$ (see \cite{lb} for other modes) is
\be
{\cal {H}}_{eff} = \gff
\left\{V_{ub}V_{us}^* a_2-2V_{tb}V_{ts}^*(a_3-a_5)\right\}
(\bar s_\alpha \gam_\mu (1-\gam_5) b_\alpha)
(\bar u_\beta \gam^\mu (1-\gam_5) u_\beta)
\ee
where of the Wilson coefficients $a_i$'s, $a_2$ comes from tree level 
operators and $a_{3,5}$ from QCD penguins \cite{buras}.
The invariant amplitude for the hadronic process $\lbb \ra \lb \pi^+
\pi^-$ is, then,
\be
M = (X+Y)q^\mu \lln \lb | \bar s \gam_\mu (1-\gam_5) b | \lbb \rln
\label{decamp}
\ee
where
\bea
X &=& \gff F^{\pi \pi}(q^2)
V_{ub}V_{us}^* a_2
\label{xx}\\
Y &=& - 2 \gff F^{\pi \pi}(q^2)V_{td}V_{ts}^*(a_3-a_5)
\label{yy}
\eea
with
$a_2 \equiv c_2^{eff}+c_1^{eff}/N_c$ and
$a_{3(5)} \equiv c_{3(5)}^{eff}+c_{4(6)}^{eff}/N_c$ and
$F^{\pi \pi}(q^2) = [1-q^2/m_{\pi \pi}^2
+i\Gamma_\sigma/m_{\pi \pi}]^{-1}$.
In the following, we take the Wilson coefficients, $a_i$, in the
limit $N_c \ra \infty$ and for numerical values we follow the
works in \cite{hyc}.
And for the form factor $F^{\pi \pi}(q^2)$, we
use $m_{\pi \pi} \approx m_\sigma = 0.7 GeV$ and $\Gamma_\sigma = 0.2
GeV$.
The matrix elements between baryons are parametrized in heavy
quark effective theory as
\be
\lln \lb | \bar s \gam_\mu (1-\gam_5) b | \lbb \rln = \bar u_\lb
[F_1 + v\hspace{-2mm}/\hspace{0.3mm} F_2]\gam_\mu (1-\gam_5)
u_{\lbb} \label{bmat} \ee where $v = p_{\lbb}/m_{\lbb}$ and $F_1 =
0.473$ and $F_2 = -0.117$ as predicted in QCD sum rules analysis
\cite{ma} that satisfy the bound \cite{gc}  $F_2/F_1 \approx
-0.25$ \cite{Pole}.
The invariant amplitude squared with all spins averaged, letting
pion mass to vanish, using eq. (\ref{bmat}), from eq.
(\ref{decamp}) is \bea |MM^\dagger| &=& {G_f^2 \over 2} [F^{\pi
\pi}(q^2)]^2
             \{|V_{ub}V_{us}^*|^2a_2^2+4|V_{tb}^*V_{ts}|^2(a_3-a_5)^2
                -4|V_{ub}V_{us}V_{tb}^*V_{ts}|a_2(a_3-a_5)\} \nonumber 
\\
&& \times [F_1^2+F_1F_2(1+m_\lb/m_{\lbb})+F_2^2]
(m_{\lbb}+m_\lb)^2(m_{\lbb}-m_\lb)^2
\eea
On performing the integration using RAMBO\footnote{We thank Prof. D. 
London for providing the programme.}, the prediction for the
branching ratio is
\be
BR(\lbb \ra \lb \pi^+ \pi^-) \equiv 4.4 \times 10^{-8} \ee
Like CP asymmetry, TP asymmetry does arise when two amplitudes
interfere, provided their weak phases are different from one
another. The size of the TPA depends on the relative magnitudes of
the coefficients $X$ and $Y$ in eqs. (\ref{xx}) and (\ref{yy}). On
looking for TPC's in this process, we obtain from eq.
(\ref{decamp}) the following 
\bea
|MM^\dagger|_1^{T-odd}
&=& 16 \xi F_2^2[(m_{\lbb}-m_\lb)^2m_{\lb} E_\pi
+(m_{\lbb}-m_\lb)m_{\lbb}m_\lb E_\pi] {\bf \hat p_\pi \cdot (\hat
s_\lb \times \hat s_{\lbb})}\\
|MM^\dagger|_2^{T-odd} &=& 3\xi
(2F_2^2-F_1F_2)(m_{\lbb}-m_\lb)^2m_{\lbb}m_\lb {\bf \hat p_\Lambda
\cdot (\hat s_\lb \times \hat s_{\lbb})}\\ 
|MM^\dagger|_3^{T-odd} &=& 12\xi
F_2^2(m_{\lbb}-m_\lb)m_{\lbb}m_\lb E_\pi {\bf \hat s_\lb \cdot
(\hat p_\pi \times \hat p_{\lb})}\\ |MM^\dagger|_4^{T-odd} &=&
-4\xi F_2^2(m_{\lbb}-m_\lb)m_{\lbb}m_\lb E_\pi {\bf \hat s_{\lbb}
\cdot (\hat p_\pi \times \hat p_\lb)} \eea where the subcripts
denote labelling the amplitude squared referring to the TPC in the
right hand side and
\be
\xi = 2G_f^2a_2(a_3-a_5)[F^{\pi
\pi}(q^2)]^2Im[V_{ub}V_{us}^*V_{tb}^*V_{ts}e^{i \delta}]
\ee
We find the magnitude of the TPA's corresponding to each of the TPC's:
\bea
{\cal {A}}^1_T = 7.2\%\\
{\cal {A}}^2_T = 6.5\%\\
{\cal {A}}^3_T = 4.3\%\\
{\cal {A}}^4_T = 1.4\%
\label{tp4}
\eea
where we have assumed $\delta=0$ and expressed the weak phase in terms 
of 
Wolfenstein parameters and
one of them $\eta = 0.35$. We can now see that the TPA corresponding to
a TPC involving two polarisation vector is larger in magnitude as
expected of than the one involving single polarisation.
In conclusion, we have found the branching ratio of the decay
$\lbb \ra \lb \pi^+ \pi^-$ and computed TPA in $\lbb \ra \lb \pi^+
\pi^-$ decays. We have earlier \cite{sa} pointed out that $\lb$
spin constituted TPC appear within the standard model. The same
is expected to hold in this decay. The present predictions for TPA 
except 
the
one in eq. (\ref{tp4}) are accessible in a collider with $10^{10}$
$\lbb \bar \lbb$ pair production.
\acknowledgements
This work is financially supported by the National Science Council of
Republic of China under the contract number NSC-91-2112-M-007-043.

\references
\bibitem{gv} G. Valencia, \pr {\bf D 39} (1989) 3339.
\bibitem{bk} B. Kayser, \np {\bf B 13} (Proc. Suppl.) (1990) 487.
\bibitem{dl} W. Bensalem and D. London \pr {\bf D 64} (2001) 116003;
W. Bensalem, A. Datta and D. London \pl {\bf B 538} (2002) 309.
\bibitem{sa} S. Arunagiri and C. Q. Geng, hep-ph/0305268.
\bibitem{dl0} W. Bensalem, A. Datta and D. London, \pr {\bf D 66} 
(2002)
094004.
\bibitem{dl1} Huge TPA in ${B \ra VV}$ decays arises in models with new
physics. See, A. Datta and D. London, hep-ph/0303159.
\bibitem{lb} J. G. Korner and M. Kramer \pl {\bf B 275} (1992) 495;
T. Mannel and S. Recksiegel, J. Phys. {\bf G 24} (1998) 979; Acta
Phys. Polon. {\bf B 28} (1997) 2489; C. H. Chen, C. Q. Geng and J.
N. Ng, \pr {\bf D 65} (2002) 091502; hep-ph/0210067.
\bibitem{buras} G. Buchalla, A. J. Buras and M. E. Lantenbatcher, Rev. 
Mod. Phys. {\bf 68},(1996) 1125. 
\bibitem{hyc} A. Ali, G. Kramer and C-D. Lu, \pr {\bf D 58} (1998) 
094009;
H. Y. Cheng and K. C. Yang, \pr {\bf D 66}
(2002) 014020.
\bibitem{ma}
C. S. Huang and H. G. Yan, \pr {\bf D 59} (1999) 114022; M. R.
Ahmady, C. S. Kim, S. Oh and C. Yu, hep-ph/0305031.
\bibitem{gc} CLEO Collaboration, G. Crawford {\em et al}, \prl {\bf 75} 
(1995) 624.
\bibitem{Pole}See also, T.~Mannel, W.~Roberts and Z.~Ryzak,
Nucl.\ Phys.\ B {\bf 355} (1991) 38; C.~H.~Chen and C.~Q.~Geng,
Phys.\ Lett.\ B {\bf 516} (2001) 327.
\end{document}